\documentclass[a4paper]{jpconf}
\usepackage{graphicx}
\newcommand{\be}{\begin{equation}}
\newcommand{\ee}{\end{equation}}
\newcommand{\bea}{\begin{eqnarray}}
\newcommand{\eea}{\end{eqnarray}}
\newcommand{\nn}{\nonumber\\}
\begin{document}
\title{Dilepton Production from Parton Interactions in the
Early Stage of Relativistic Heavy-Ion Collisions}

\author{O. Linnyk$^a$, E. L. Bratkovskaya$^a$, J. Manninen$^b$,
and W. Cassing$^b$}

\address{$^a$Institut f\"ur Theoretische Physik, Universit\"{a}t Frankfurt, 60438 Frankfurt am Main,
  Germany \\
  $^b$Institut f\"ur Theoretische Physik, Universit\"at Giessen,
  35392 Giessen, Germany}

\ead{linnyk@fias.uni-frankfurt.de}

\begin{abstract}
We address the dilepton production from the parton interactions in
the early stage of relativistic heavy-ion collisions within the
parton-hadron-string dynamics (PHSD) off-shell transport approach.
The description of partons in PHSD is based on the dynamical
quasiparticle model (DQPM) matched to reproduce lattice QCD results
in thermodynamic equilibrium. According to the DQPM the constituents
of the strongly interacting quark-gluon plasma (sQGP) are massive
and off-shell quasi-particles (quarks and gluons) with broad
spectral functions. In order to address the electromagnetic
radiation of the sQGP, we derive off-shell cross sections of $q\bar
q\to\gamma^*$, $q\bar q\to\gamma^*+g$ and $qg\to\gamma^*q$ ($\bar q
g\to\gamma^* \bar q$) reactions taking into account the effective
propagators for quarks and gluons from the DQPM. Dilepton production
in In+In collisions at 158 AGeV and in Au+Au at $\sqrt{s}=200$ GeV
is calculated by implementing these off-shell processes into the
PHSD transport approach. By comparing to the data of the NA60 and
PHENIX Collaborations, we study the relative importance of different
dilepton production mechanisms and point out the regions in phase
space where partonic channels are dominant.
\end{abstract}

\vspace{-0.3pt}

Dileptons are emitted over the entire space-time evolution of the
heavy-ion collision, from the initial nucleon-nucleon collisions
through the hot and dense phase and to the hadron decays after the
freeze-out. This is both a challenge and advantage of the probe. The
separation of different ``physics" in the dilepton radiation is
nontrivial due to the non-equilibrium nature of the heavy-ion
reactions, and covariant transport models have to be used to
disentangle the various sources that contribute to the final
dilepton spectra seen experimentally.

\vspace{-0.3cm}
\section{The PHSD approach to dilepton radiation}

To address the dilepton production in a hot and dense medium -- as
created in heavy-ion collisions -- we employ an up-to-date
relativistic transport model, the Parton Hadron String
Dynamics~\cite{CasBrat} (PHSD), that incorporates the explicit
partonic phase in the early reaction region. Within PHSD, one solves
generalized transport equations on the basis of the off-shell
Kadanoff-Baym equations for Green's functions in phase-space
representation (in the first order gradient expansion beyond the
quasiparticle approximation).  The approach consistently describes
the full evolution of a relativistic heavy-ion collision from the
initial hard scatterings and string formation through the dynamical
deconfinement phase transition to the quark-gluon plasma (QGP) as
well as hadronization and to the subsequent interactions in the
hadronic phase.
In the hadronic sector PHSD is equivalent to the
Hadron-String-Dynamics (HSD) transport approach
\cite{CBRep98,Brat97,Ehehalt} that has been used for the description
of $pA$ and $AA$ collisions from SIS to RHIC energies. In
particular, (P)HSD incorporates off-shell dynamics for vector mesons
-- according to Refs.~\cite{Cass_off1} -- and a set of vector-meson
spectral functions~\cite{Brat08} that covers possible scenarios for
their in-medium modification.
%

In the {\it off-shell} transport description, the hadron spectral
functions change dynamically during the propagation through the
medium and evolve towards the on-shell spectral function in the
vacuum.
As demonstrated in~\cite{Brat08} the off-shell dynamics is important
for resonances with a rather long lifetime in vacuum but strongly
decreasing lifetime in the nuclear medium (especially $\omega$ and
$\phi$ mesons) and also proves vital for the correct description of
dilepton decays  of $\rho$ mesons with masses close to the two pion
decay threshold.

Let us remind the off-shell transport equations
(see~\cite{shladming} for details). One starts with a first order
gradient expansion of the Wigner transformed Kadanoff-Baym equation
using the following expression for the memory integrals
\bea \label{eq:i1csnew} I_1^{><}(x_1,x_2) =& -
\int_{-\infty}^{\infty} d^{D}x^{\prime}~ \Theta(t_1-t^{\prime})~
     \left[ \Sigma^{>}(x_1,x^{\prime}) - \Sigma^{<}(x_1,x^{\prime}) \right]~
     G^{><}(x^{\prime},x_2) \\[0.2cm]
 & + \int_{-\infty}^{\infty} d^{D}x^{\prime} ~ \Sigma^{><}(x_1,x^{\prime})~
   \Theta(t_2-t^{\prime}) ~ \left[ G^{>}(x^{\prime},x_2) - G^{<}(x^{\prime},x_2) \right] \nn
=& - \int_{-\infty}^{\infty} d^{D}x^{\prime} ~
\Sigma^{R}(x_1,x^{\prime}) ~
   G^{><}(x^{\prime},x_2) ~+~ \Sigma^{><}(x_1,x^{\prime}) ~ G^{A}(x^{\prime},x_2)~. {}
\eea
and arrives at a generalized transport equation
\cite{Cass_off1,other_off}:
%
\bea \label{eq:general_transport} \hspace{-3em}\underbrace{ 2
p^{\mu}
\partial^{x}_{~\mu} i~\bar{G}^{><}
    - \{ \bar{\Sigma}^{\delta} + Re~\bar{\Sigma}^{R}, i\bar{G}^{><} \} }
& - \{ i~\bar{\Sigma}^{><},~ Re~\bar{G}^{R} \}
  = i~\bar{\Sigma}^{<} ~ i\bar{G}^{>} - i~\bar{\Sigma}^{>} ~ i~\bar{G}^{<} \\[0.2cm]
 \{ \bar{M},~ i ~ \bar{G}^{><} \} \hspace{4.5em}
& - \{ i~\bar{\Sigma}^{><},~ Re~\bar{G}^{R} \}
  = i~\bar{\Sigma}^{<} ~ i\bar{G}^{>} - i~\bar{\Sigma}^{>} ~ i~\bar{G}^{<}  {}
\eea
as well as a generalized mass-shell equation
\bea \label{eq:general_mass} \hspace{-2em}\underbrace{ [ p^2 - m^2 -
\bar{\Sigma}^{\delta} - Re~\bar{\Sigma}^{R} ]}_{\bar{M}}
i\bar{G}^{><} = i\bar{\Sigma}^{><} Re~\bar{G}^{R} + \frac{1}{4}~
\{i\bar{\Sigma}^{>}, i\bar{G}^{<} \} - \frac{1}{4}~
\{i\bar{\Sigma}^{<}, i\bar{G}^{>} \} \eea
with the mass-function $\bar{M}$.
In the transport equation (\ref{eq:general_transport}) one
recognizes on the l.h.s. the drift term $p^{\mu}~\partial^{x}_{~\mu}
~ i\bar{G}^{><}$, as well as the Vlasov term with the local
self-energy $\bar{\Sigma}^{\delta}$ and the real part of the
retarded self-energy $Re~\bar{\Sigma}^{R}$. On the other hand the
r.h.s. represents the collision term with its typical `gain and
loss' structure.

The mass-function for fermions is
\bea \label{massfunction3} {M}_F(p,x) ~=~  \Pi_0^2 - \vec{\Pi}^{~2}
- m_h^{*2} ~ , \eea
with the effective mass and four-momentum given by
\bea \label{Ehg26}
m_h^* (x,p)&~=~ m_h + U_h^{{S}}(x,p), \mbox{  } 
\Pi^\mu (x,p)&~=~ p^\mu-U^\mu_h (x,p)~,  {} \eea
where  $U_h^S(x,p)$ and $U_h^\mu(x,p)$ denote the real part of the
scalar and vector self-energies of the particle, respectively, and
$m_h$ stands for its the bare (vacuum) mass.
After inserting (\ref{massfunction3}) into the generalized transport
equation (\ref{eq:general_transport}), the covariant off-shell
transport theory emerges, that has been denoted as
HSD~\cite{CBRep98,Ehehalt}. It is formally written as a coupled set
of transport equations for the phase-space distributions $N_h(x,p)$
$(x=(t,\vec r), ~ p=(\varepsilon,\vec p))$ of fermion $h$
\cite{CBRep98,Ehehalt} with a spectral function $A_h(x,p)$ (using
$i\bar{G}^{<}_h(x,p)=N_h(x,p) A_h(x,p)$), i.e.~\cite{voka}
\bea & \left\{ \left( \Pi_\mu-\Pi_\nu\partial_\mu^p U_h^\nu
  - m_h^*\partial^p_\mu U_h^S \right)\partial_x^\mu
  + \left( \Pi_\nu \partial^x_\mu U^\nu_h + m^*_h \partial^x_\mu U^S_h\right) \partial^\mu_p \right\}
    N_h(x,p) ~ A_h(x,p)                          \nn
&  -\{ i{\Sigma}^{<},  Re~{G}^{R} \}
 = \sum_{h_2 h_3 h_4} Tr_2 Tr_3 Tr_4 ~
   [T^{\dagger} T]_{12\rightarrow 34}
   \delta^4 (\Pi +\Pi_2-\Pi_3-\Pi_4 ) \label{Ehg24} \\
& \hspace*{15em}  \times  A_{h}(x,p) A_{h_2}(x,p_2) A_{h_3}(x,p_3)
A_{h_4}(x,p_4) \nn & \times \left\{ N_{h_3}(x,p_3) N_{h_4}(x,p_4)
\bar{f}_h(x,p)
   \bar{f}_{h_2}(x,p_2) - N_h(x,p) N_{h_2}(x,p_2)\bar{f}_{h_3}(x,p_3)
   \bar{f}_{h_4}(x,p_4) \right\}.  {}
\eea
Here $\partial^x_\mu \equiv (\partial_t, \vec\nabla_r)$ and
$\partial^p_\mu \equiv (\partial_\varepsilon, \vec\nabla_p),
(\mu=0,1,2,3)$. The backflow term in (\ref{Ehg24}) is given by
\bea \label{eq:backflow}
 -\{ i{\Sigma}^{<}, Re{G}^{R} \}
& = \partial^\mu_p \left(\frac{ M_h(x,p)}{ M_h(x,p)^2
  + \Gamma_h(x,p)^2/4}\right) ~ \partial_\mu^x
         \left(N_h(x,p) ~ \Gamma_h(x,p) \right)  \\[0.2cm]
& - \partial_\mu^x \left(\frac{ M_h(x,p)}{ M_h(x,p)^2
  + \Gamma_h(x,p)^2/4}\right) ~ \partial^\mu_p
         \left( N_h(x,p) ~ \Gamma_h(x,p) \right).  {}
\eea
It stands for the off-shell evolution which vanishes in the on-shell
limit, when the spectral function $A_h(x,p)$ does not change its
shape during the propagation through the medium, i.e. ${\vec
\nabla}_r \Gamma(x,p)$=0 and ${\vec \nabla}_p \Gamma(x,p)$=0. The
set of coupled differential-integral equations (\ref{Ehg24}) is
solved using a test-particle ansatz.

The transport description of quarks and gluons in PHSD is based on
the knowledge of the partonic propagators from a dynamical
quasiparticle model for partons matched to reproduce lattice QCD
results in thermodynamic equilibrium (DQPM). The DQPM describes QCD
properties in terms of single-particle Green's functions (in the
sense of a two-particle irreducible approach) and leads to the
notion of the constituents of the sQGP being effective
quasiparticles, which are massive and have broad spectral functions
(due to large interaction rates). The transition from partonic to
hadronic degrees of freedom in PHSD is described by covariant
transition rates for the fusion of quark-antiquark pairs to mesonic
resonances or three quarks (antiquarks) to baryonic states.

\begin{figure}
\centering
\includegraphics[width=0.9\textwidth]{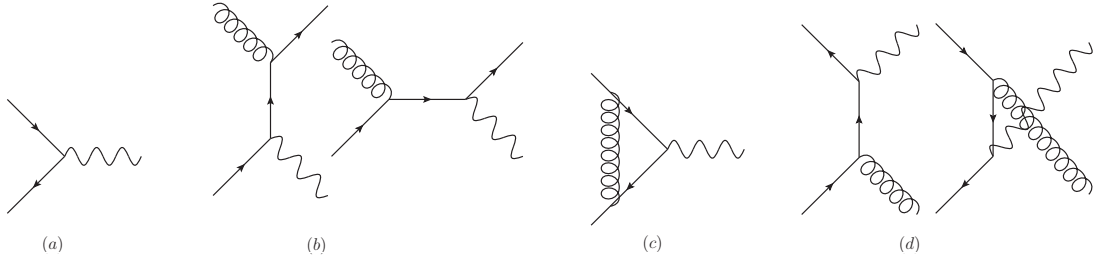}
\caption{Diagrams contributing to the dilepton production from a
QGP: (a) Born mechanism, (b) gluon-Compton scattering (GCS), (c)
vertex correction, (d) gluon Bremsstrahlung (NLODY), where virtual
photons (wavy lines) split into lepton pairs; spiral lines denote
gluons, arrows denote quarks. In each diagram the time runs from
left to right.} \label{diagrams}
\end{figure}

Dilepton radiation by the constituents of the QGP proceeds via the
elementary processes illustrated in Fig.~\ref{diagrams}: the Born
(Drell-Yan) $q+\bar q$ annihilation mechanism, Gluon Compton
scattering ($q+g\to \gamma^*+q$ and $\bar q+g\to \gamma^*+\bar q$),
and quark + anti-quark annihilation with gluon Bremsstrahlung in the
final state ($q+\bar q\to g+\gamma^*$). In the on-shell
approximation, one uses perturbative QCD cross sections for the
processes listed above.
However, in order to make quantitative predictions at experimentally
relevant low dilepton mass and strong coupling, we have to take into
account the non-perturbative spectral functions and self-energies of
the quarks, anti-quarks and gluons thus going beyond the on-shell
approximation.
For this purpose, off-shell cross sections were derived
in~\cite{olena2010} for dilepton production in the reactions $q+\bar
q\to l^+l^-$ (Drell-Yan mechanism), $q+ \bar q\to g+l^+l^-$ (quark
annihilation with the gluon Bremsstrahlung in the final state),
$q(\bar q)+g\to q(\bar q)+ l^+l^-$ (gluon Compton scattering), $g\to
q+\bar q+l^+l^-$ and $q(\bar q)\to q(\bar q)+g+l^+l^-$ (virtual
gluon decay, virtual quark decay) in the sQGP in effective
perturbation theory by dressing the quark and gluon lines with the
DQPM propagators for quarks and gluons. The obtained off-shell
elementary cross sections then are implemented into the PHSD
transport code, where the masses of quarks and gluons are
distributed according to the DQPM spectral functions.

\begin{figure*}
  \begin{minipage}[b]{0.49\textwidth}
    \includegraphics[width=.95\textwidth]{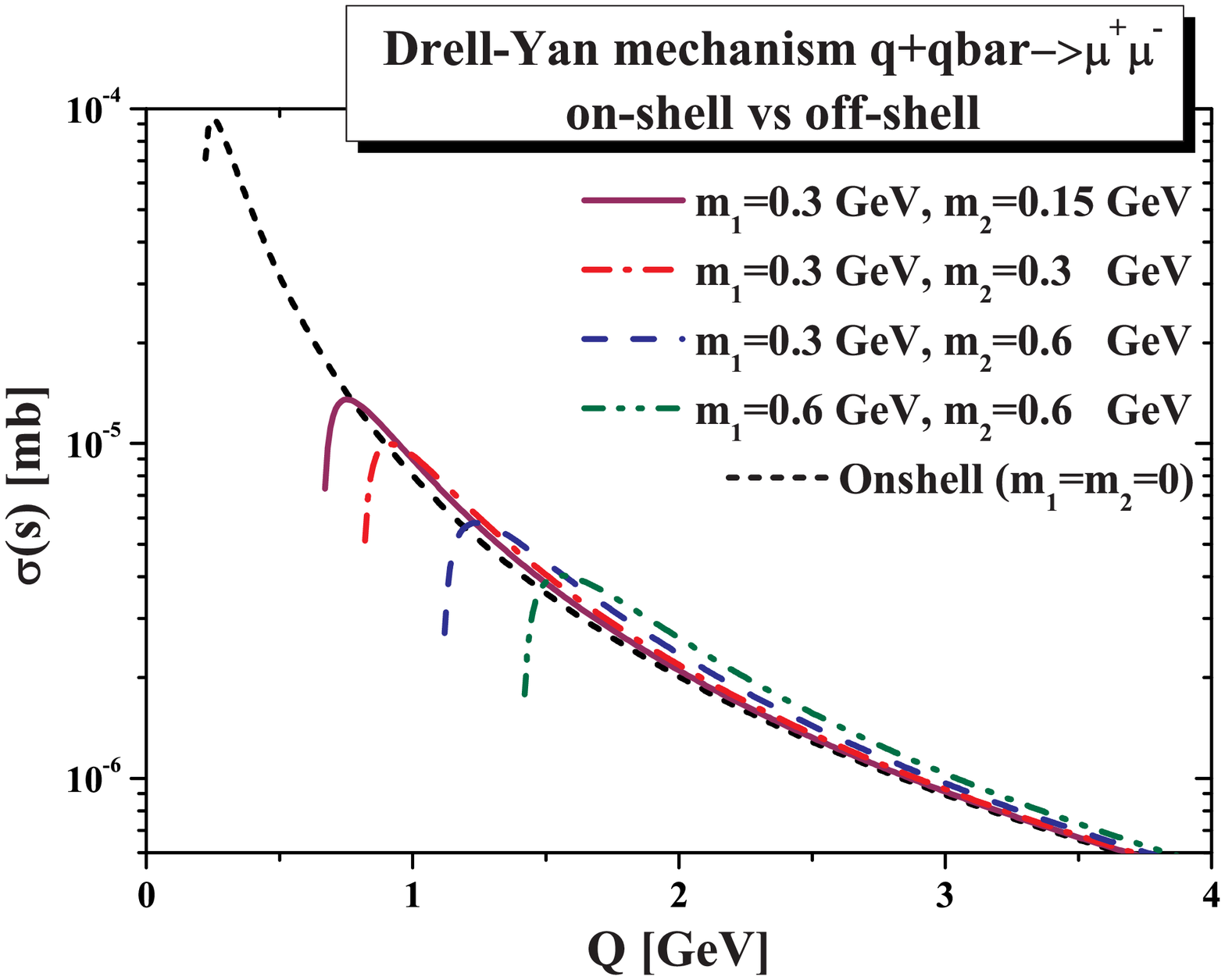}
    \caption{Dilepton production cross sections in the
Born channel ($q+\bar q\to \mu^+ + \mu^-$). The short dashed (black)
line shows the on-shell, i.e. the standard pQCD result. The other
lines show the off-shell cross section, in which the annihilating
quark and antiquark have finite masses $m_1$ and $m_2$ with
different values.}
    \label{DYoffVsOn}
  \end{minipage}
  \hspace{0.02\textwidth}
  \begin{minipage}[b]{0.49\textwidth}
    \includegraphics[width=\textwidth]{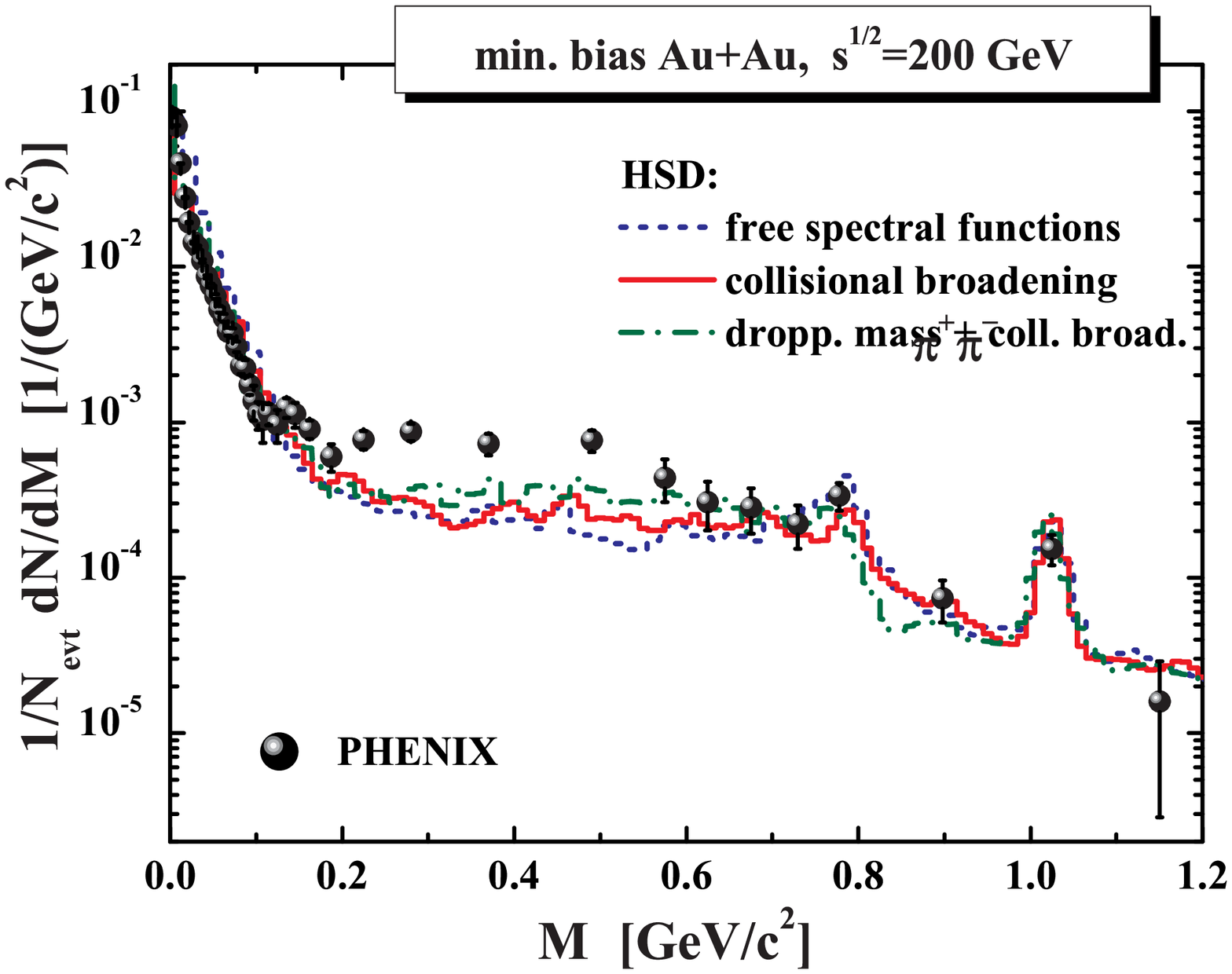}
     \caption{ The HSD
results  for the mass differential dilepton spectra in case of
inclusive $Au + Au$ collisions at $\sqrt{s}$ = 200 GeV in comparison
to the data from PHENIX~\protect{\cite{PHENIX,PHENIXlast}} in the
`free' scenario, the `collisional broadening' picture as well as the
`dropping mass + collisional broadening' model.} \label{HSD2}
  \end{minipage}
 \end{figure*}

For illustration, we plot the dilepton production cross sections in
the Born mechanism in Fig.~\ref{DYoffVsOn}. The short dashes (black)
line shows the on-shell, i.e. the standard perturbative result. The
other lines show the off-shell cross section, in which the
annihilating quark and antiquark have finite masses $m_1$ and $m_2$
with different values: $m_1=0.3$~GeV, $m_2=0.15$~GeV (solid magenta
line), $m_1=0.3$~GeV, $m_2=0.3$~GeV (dash-dotted red line),
$m_1=0.3$~GeV, $m_2=0.6$~GeV (dashed blue line), $m_1=0.6$~GeV,
$m_2=0.6$~GeV (dash-dot-dot green line).

\vspace{-0.3cm}
\section{Comparison to data}

By employing the HSD approach to the low mass dilepton production in
relativistic heavy-ion collisions, it was shown in~\cite{here} that
the NA60 Collaboration data for the invariant mass spectra for
$\mu^+\mu^-$ pairs from In+In collisions at 158 A$\cdot$GeV favored
the `melting $\rho$' scenario \cite{NA60}. On the other hand, the
dilepton spectrum from In+In collisions at 158 A$\cdot$GeV for
$M>1$~GeV could not be accounted for by the known hadronic sources
(see Fig.2 of~\cite{here}). Also, hadronic models do not reproduce
the softening of the $m_T$ distribution of dileptons at
$M>1$~GeV~\cite{NA60}.

In Refs.~\cite{Linnyk:2009nx,Linnyk:2010ar,HPproceedings} we
presented PHSD results for the dilepton spectrum as produced in
$In+In$ reactions at 158~AGeV compared to the NA60
data~\cite{NA60,Arnaldi:2008er}. We confirmed the HSD results that
the spectrum at invariant masses below 1~GeV was better reproduced
by the $\rho$ meson yield, if a broadening of the meson spectral
function in the medium was assumed. On the other hand, the spectrum
at $M>1$~GeV was shown to be dominated by the partonic sources.
Moreover, accounting for partonic dilepton sources allowed to
reproduce in PHSD the effective temperature of the dileptons (slope
parameters) in the intermediate mass range. The softening of the
transverse mass spectrum with growing invariant mass implies that
the partonic channels occur dominantly before the collective radial
flow has developed.

The PHENIX Collaboration has presented dilepton data from $pp$ and
$Au+Au$ collisions at Relativistic-Heavy-Ion-Collider (RHIC)
energies of $\sqrt{s}$=200~GeV~\cite{PHENIXpp,PHENIX,PHENIXlast},
which show a large enhancement in $Au+Au$ reactions (relative to
scaled $pp$ collisions) in the invariant mass regime from 0.15 to
0.6 GeV/c$^2$. Moreover, if realistic partial loss of the $D$- and
$\bar D$-meson correlations due to their rescattering is kept in
mind, one has to conclude that there exists another domain of
invariant mass, in which the measured dilepton yield in $A+A$
collisions is underestimated by the scaled yield from $p+p$; this
domain is at masses from 1 to
4~GeV/c$^2$~\cite{Linnyk:2010ar,Manninen:2010yf}.

Let us first present in Fig~\ref{HSD2} the HSD results for $e^+e^-$
pairs in inclusive $Au + Au$ collisions in comparison to the data
from PHENIX~\cite{PHENIX,PHENIXlast} as calculated in~\cite{here},
recalling that HSD provides a reasonable description of hadron
production in $Au+Au$ collisions at $\sqrt{s}$ = 200
GeV~\cite{Brat03}.
Whereas the total yield  is quite well described in the region of
the pion Dalitz decay as well as around the $\omega$, $\phi$ and
$J/\Psi$ mass, HSD clearly underestimates the measured spectra in
the regime from 0.2 to 0.6 GeV by approximately a factor of
5~\cite{here}. After including the in-medium modification scenarios
for the vector mesons, we achieve a sum spectrum which is only
slightly enhanced compared to the 'free' scenario (see Fig.~5).  The
low mass dilepton spectra from $Au+Au$ collisions at RHIC (from the
PHENIX Collaboration) are clearly underestimated in the invariant
mass range from 0.2 to 0.6 GeV in the 'collisional broadening'
scenario as well as in the 'dropping mass + collisional broadening'
model. We mention that HSD results for the low mass dileptons are
very close to the calculated spectra from van Hees and Rapp as well
as Dusling and Zahed~\cite{Dussi} (cf. the comparison in
Ref.~\cite{PHENIXlast,AToia}).
At higher masses (from 1 to 4 GeV) the only hadronic sources of
correlated lepton pairs are the charmed mesons: semi-leptonic decays
of correlated D-mesons and the dilepton decays of charmonia. Between
the $\phi$ and $J/\Psi$ peaks, the HSD results underestimate the
PHENIX data by approximately a factor of two.

\begin{figure*}
  \begin{minipage}[b]{0.49\textwidth}
    \includegraphics[width=\textwidth]{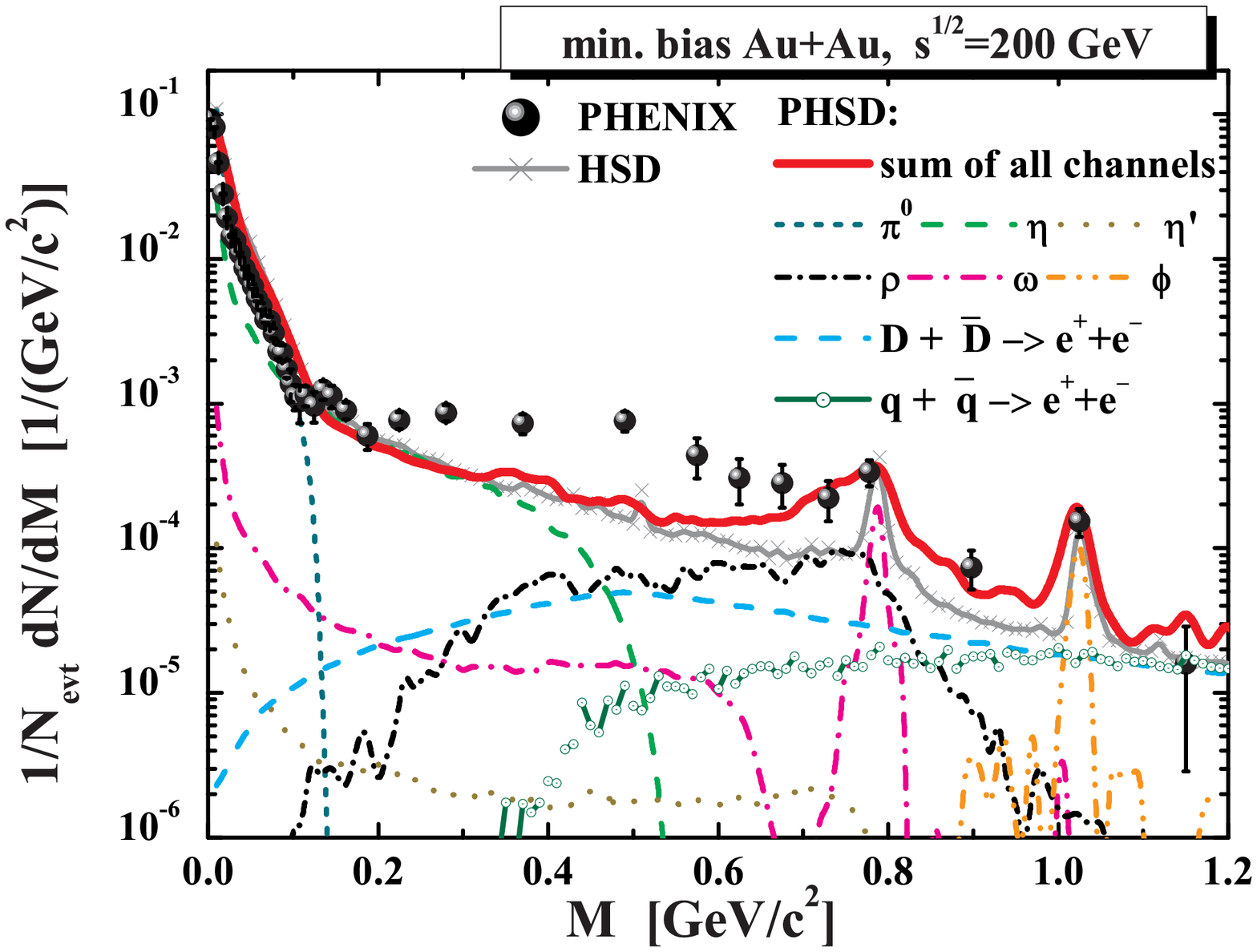}
    \caption{The PHSD results  for the mass differential dilepton spectra in case of
inclusive $Au + Au$ collisions at $\sqrt{s}$ = 200 GeV in comparison
to the data from PHENIX~\protect{\cite{PHENIX,PHENIXlast}} in the
low mass region ($M \! = \! 0 \! - \! 1.2$~GeV). } \label{PHSD1}
  \end{minipage}
  \hspace{0.02\textwidth}
  \begin{minipage}[b]{0.49\textwidth}
    \includegraphics[width=\textwidth]{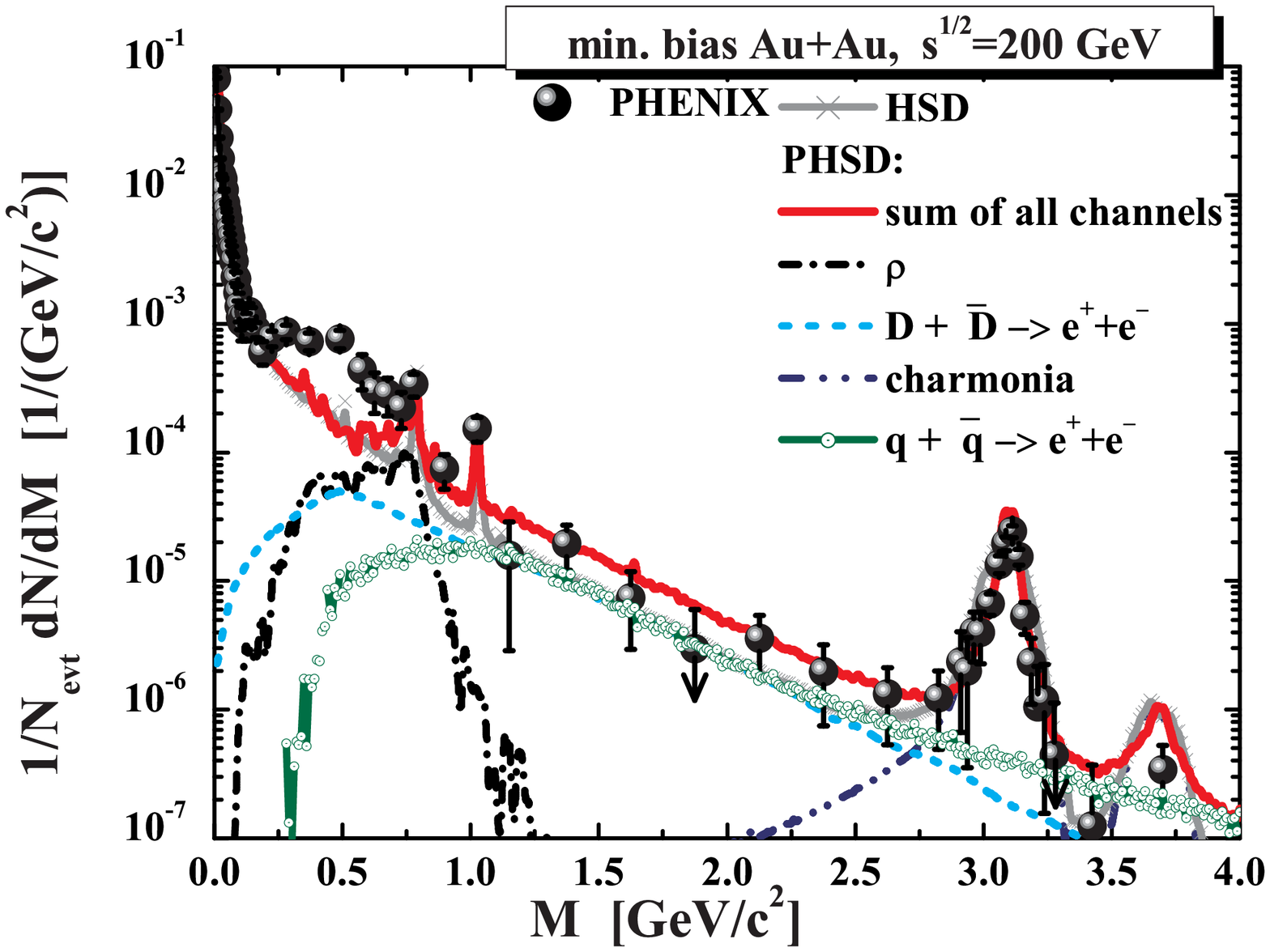}
     \caption{ The PHSD
results  for the mass differential dilepton spectra in case of
inclusive $Au + Au$ collisions at $\sqrt{s}$ = 200 GeV in comparison
to the data from PHENIX~\protect{\cite{PHENIX,PHENIXlast}} for
$M\!=\!0\!\,-\,\!4$~GeV. } \label{PHSD2}
  \end{minipage}
\end{figure*}

By implementing the off-shell partonic processes into the PHSD
transport approach, we calculate the dilepton spectra in $Au+Au$ at
$\sqrt{s}$=200~GeV and compare to the PHENIX data in
Figs.~\ref{PHSD1} and \ref{PHSD2}. In Fig.~\ref{PHSD1} we show our
results for low masses ($M=0-1.2$~GeV); in this region, the yield in
PHSD is dominated by hadronic sources and essentially coincides with
the HSD result. Note that the collisional broadening scenario for
the modification of the $\rho$-meson was used in the calculations
presented in the Figs.~\ref{PHSD1} and \ref{PHSD2}. There is a
discrepancy between the PHSD calculations and the data in the region
of masses from 0.2 to 0.6 GeV. The discrepancy is not amended by
accounting for the radiation from the QGP, since the latter is
`over-shone' by the radiation from hadrons integrated over the
evolution of the collision. In contrast, the partonic radiation is
considerable in the mass region $M=1-4$~GeV as seen in
Fig.~\ref{PHSD2}. The dileptons generated by the quark-antiquark
annihilation in the sQGP constitute about half of the observed yield
in the mass range between the masses of the $\phi$ and the $J/\Psi$
mesons. For $M>2.5$~GeV the partonic yield dominates over the
D-meson contribution. Thus, accounting for partonic radiation in
PHSD fills up the gap between the hadronic model
results~\cite{here,Manninen:2010yf} and the data at $M>1$~GeV.

\vspace{-0.3cm}
\section{Summary}

The Parton Hadron String Dynamics~\cite{CasBrat} (PHSD) transport
approach incorporates the relevant off-shell dynamics of the vector
mesons as well as the explicit partonic phase in the early hot and
dense reaction region.
A comparison of the transport calculations to the data of the NA60
Collaborations points towards a 'melting' of the $\rho$-meson at
high densities, i.e. a broadening of the vector meson's spectral
function. On the other hand, the spectrum for $M>1$~GeV is shown to
be dominated by the partonic sources.

The low mass dilepton spectra from $Au+Au$ collisions at RHIC (from
the PHENIX Collaboration) are clearly underestimated by the hadronic
channels in the invariant mass range from 0.2 to 0.6 GeV. The
discrepancy is not amended by accounting for the radiation from the
QGP, since the latter is `over-shone' by the radiation from hadrons
integrated over the evolution of the collision.
In contrast, the partonic radiation is visible in the mass region
$M=1-4$~GeV. The dileptons generated by the quark-antiquark
annihilation in the sQGP constitute about half of the observed yield
in the mass range between the masses of the $\phi$ and the $J/\Psi$
mesons. For $M>2.5$~GeV the partonic yield dominates over the
D-meson contribution. Thus, accounting for partonic radiation in
PHSD fills up the gap between the hadronic model results and the
data for $M>1$~GeV.


\hspace{-0.65cm} Work supported in part by the "HIC for FAIR"
framework of the "LOEWE" program and by DFG. We acknowledge
stimulating discussions with S. Damjanovic, V. Konchakovski and A.
Toia.

\end{document}